\documentclass[multphys,vecphys]{svmult}

\usepackage{makeidx}
\usepackage{graphicx}
\usepackage{multicol}

\makeindex

\begin{document}

\title*{Structure and Stability of Prestellar Cores}

\author{Jens Kauffmann\and
Frank Bertoldi}

\institute{Max-Planck-Institut f\"ur Radioastronomie, Auf dem H\"ugel 69,
D-53121 Bonn, \texttt{jkauffma@mpifr-bonn.mpg.de} and
\texttt{bertoldi@mpifr-bonn.mpg.de}}

\maketitle

\section*{Abstract}

Following an approach initially outlined by McKee \& Holliman \cite{mckee1999},
we investigate the structure and stability of dense, starless molecular
cloud cores. We model those as spherical clouds in hydrostatic
equilibrium and supported against gravity by thermal, turbulent, and magnetic
pressure. We determine the gas pressure by solving for thermal equilibrium between heating and
cooling, while the
turbulent and magnetic pressures are assumed to obey polytropic
equations of state.
In comparing the models to observed cloud cores we find that the
observed peak column densities often exceed the limit for stable
equilibria supported by thermal pressure alone, suggesting significant
non-thermal pressure if the cores are to be
stable. Non-thermal support is also needed to stabilize cores embedded
in molecular clouds with high average pressures.
Since the observed molecular linewidths of cores suggest
that the turbulent pressure is lower than the thermal pressure,
magnetic field are likely a dominant pressure component
in many such cores.

\section{Hydrostatic equilibrium models}

To model a gas cloud in hydrostatic equilibrium we make the simplified
assumption that the gas pressure consists of the sum
of thermal (subscript ``th''), turbulent wave (``w''),
and magnetic (``m'') pressure components,
\begin{equation}
	P = P_{\rm th} + P_{\rm w} + P_{\rm m} \, .
\end{equation}
We compute the thermal pressure through a detailed thermal equilibrium
calculation, but the wave and magnetic pressures are assumed to obey a
polytropic equation of state,
\begin{equation}
	P_{\rm w} \propto \varrho^{\gamma_{P, \rm w}} ~ ,
		\quad 
	P_{\rm m} \propto \varrho^{\gamma_{P, \rm m}} ~ ,
	\label{eq:eos_polytr}
\end{equation}
i.e. the pressure only depends on density, and $\gamma_{P}$, the
polytropic exponent, is constant in a given object.
Given the thermal and non-thermal gas pressure, it is possible to construct
hydrostatic equilibria for any surface pressure and central to surface
density contrast. We assume spherical symmetry for simplicity.

Whether a given hydrostatic equilibrium model cloud is gravitationally
stable depends on its response to perturbations in pressure or density.
Different equations of state apply for such
perturbations. For example, if the perturbation occurred on a time
scale shorter than the cooling time, the scaling of the thermal
pressure with density would be stiffer than if the perturbation would
be allowed to reach thermal equilibrium.
To analyze the stability we
make the simple assumption that the perturbation obeys a 
polytropic equation of state, but with some different ``adiabatic
index'', $\gamma$:
\begin{equation}
	\udelta P_{\rm w}/P_{\rm w,0} \propto
		(\udelta \varrho/\varrho_0)^{\gamma_{\rm w}} ~ ,
		\quad 
	\udelta P_{\rm m}/P_{\rm m,0} \propto
		(\udelta \varrho/\varrho_0)^{\gamma_{\rm m}} ~ ,
 \label{eq:eos_adiabat}
\end{equation}
where  $\udelta$ are infinitesimal
perturbations, and the subscript ``0'' refers to values before the perturbation.

A hydrostatic equilibrium cloud is
stable against spontaneous contraction or expansion when at any point in the cloud,
the pressure 
increases during a compression or decreases during an expansion, i.e.,
\begin{equation}
	\udelta P / \udelta r < 0
\end{equation}
where $r$ is the radius from the cloud center.
Marginally stable equilibria have $\udelta P / \udelta r = 0$ at some
point in the cloud.

\begin{figure}
\centering
\includegraphics[width=\linewidth]{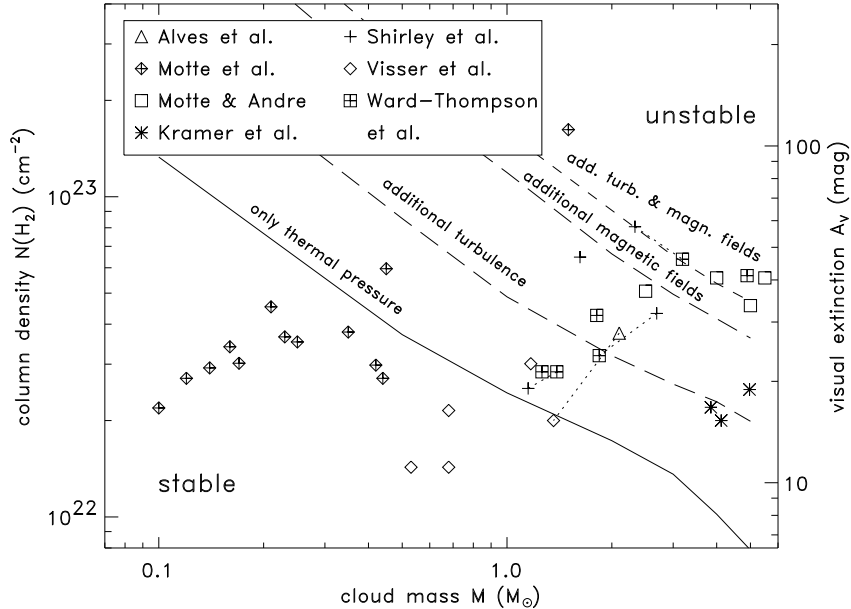}
\caption{\label{fig:1} Observed column densities of prestellar cores
(\emph{symbols}) and upper limits for stable equilibria
(\emph{lines}). The
lower line applies for pure thermal support, the one above for an
equal thermal and magnetic pressure in the cloud center, the one above that
for equal thermal and turbulent, the top one for equal thermal,
magnetic and turbulent pressures in the cloud center.
\emph{Dotted lines} connect independent measurements of the same object by
different authors.}
\end{figure}

\subsection{Thermal Pressure\label{sec:model_th}}

At any radial point in the cloud we solve the coupled thermal balance
for the gas and dust. We adopt the gas heating and cooling rates and thermal gas-dust coupling from
Goldsmith \cite{goldsmith2001}, and the dust
heating and cooling rates for the solar neighborhood from Zucconi et al. \cite{zucconi2001}. The
heating is 
due to cosmic rays and the interstellar radiation field, for which we
allow some shielding, $A_{V,0}$, due to the material surrounding the
core. When analyzing the stability, we assume that the heating or
cooling due to a perturbation is negligible. The heating and cooling rates themselves however
change, as they depend on the perturbed density and visual extinction.

\subsection{Turbulent and Magnetic Pressure\label{sec:model_nt}}

We treat the kinetic pressure due to non-thermal
motions, which we call turbulence, in the very simplifying framework of Alfv\'en waves.
The pressure due to Alfv\'en waves
can be described by a
polytropic exponent $\gamma_{P, \rm w} = 1/2$, and an adiabatic
exponent $\gamma_{\rm w} = 3/2$ for the case where no energy flow
occurs \cite{mckee1995, gammie1996}. If we allow for a flow of the turbulent energy
between the core and its surroundings during perturbations, then the adiabatic exponent would be
equal to the polytropic one, i.e. $\gamma_{\rm w} = 1/2$. This requires that
the heat flows are faster than the duration of the perturbation.

The magnetic pressure due to the average magnetic flux threading the cloud depends on the
distribution of field lines in the cloud, which is a result of how the cloud has formed while the
gas was frozen to the field lines.
A somewhat more realistic case was adopted for the 2-dimensional
hydrostatic equilibria computed by Tomisaka et al.\
\cite{tomisaka1988}: here a cloud was assumed to initially have been
spherical and at uniform density within a uniform magnetic field.
We examined these 2-D
equilibria to determine the magnetic field -- density scaling, finding
$\gamma_{P,\rm m} = 0.9$. The magnetic adiabatic exponent adopted in our work,
$\gamma_{\rm m} = 1.2$, is chosen such that the critical density contrasts of Tomisaka et al.\
\cite{tomisaka1988} are reproduced. We adopt these values for our idealizing spherical model to
study the behavior of 2-D magnetized equilibria.

\section{Results \& Conclusion}

In our yet preliminary study we have looked
at clouds with a particular ratio between the non-thermal and thermal
pressure {\it at the cloud center:} we investigated clouds with 
$(P_{\rm w}/ P_{\rm th})_{r=0}=0$ or 1, and with 
$(P_{\rm m}/ P_{\rm th})_{r=0}=0$ or 1,
which is meant to give a first impression on the effect of non-thermal
pressure support on the cloud stability. We adopt an extinction due to material surrounding the
clouds of 5 magnitudes in the visual.

As a function of cloud mass we computed the 
maximum central column density for a stable hydrostatic equilibrium.
These are plotted in Figure \ref{fig:1} for clouds with and without
non-thermal pressure. A comparison with the column densities measured
for all starless low-mass cores we found in the literature shows that
some of the more massive cores have column densities which are larger
than what could be accounted for by a stable cloud with pure thermal
pressure support.

With a magnetic or turbulent pressure at least comparable to the
thermal pressure in the cloud center, higher central condensations
comparable to those observed would be possible for stable clouds.  If
this non-thermal pressure were due to turbulent motions, then the
observed cores' non-thermal velocity dispersions should be comparable
to the sound speed, but observations typically show smaller
non-thermal linewidths \cite{caselli2002}. Thus turbulence is unlikely to be
the dominant support in such cores, and magnetic fields remain as the
stabilizing force.

A second argument for the need of non-thermal support in the more
massive low-mass cores derives from their maximum possible surface
pressure. A core of given mass in a stable hydrostatic equilibrium
can only exist if the kinetic surface pressure due to thermal and turbulent particle motions is
below some critical value. Significant non-thermal pressure is necessary to stabilize the more
massive cores ($M > {2 ~ \rm to ~ 3} \, M_\odot$) in such environments.

To summarize, our preliminary analysis of the stability of dense,
self-gravitating cores and our comparison with observed cores and
their environments suggests the need for non-thermal pressure support,
most likely provided by magnetic fields of less than $100 ~ \rm \umu G$.

\printindex
\end{document}